\documentclass[entropy,conferenceproceedings,accept,moreauthors,pdftex]{Definitions/mdpi}

\firstpage{1}
\pubvolume{0}
\issuenum{0}
\articlenumber{0}
\pubyear{2022}
\copyrightyear{2022}
\history{Received: June 2022; Accepted: -; Published: -}

\usepackage{color,xspace}
\usepackage{Definitions/macros_gpi}

\graphicspath{{Figs/}}

\setitemize{parsep=6pt,itemsep=0pt,leftmargin=*,labelsep=5.5mm}
\setenumerate{parsep=6pt,itemsep=0pt,leftmargin=*,labelsep=5.5mm} 
\setlist[description]{itemsep=0mm}   
\usepackage{environ}
\NewEnviron{myequation}{%
\begin{equation} 
\scalebox{0.95}{$\BODY$}
\end{equation}
}

\Title{Bayesian statistics approach to imaging of aperture synthesis data: RESOLVE meets ALMA.$^\dagger$}

\Author{Lukasz Tychoniec$^{1,*}$, Fabrizia Guglielmetti$^{1}$, Philipp Arras$^{2}$, Torsten En{\ss}lin$^{2}$, Eric Villard$^{1}$} 

\AuthorNames{Lukasz Tychoniec, Fabrizia Guglielmetti, Philipp Arras, Torsten En{\ss}lin, Eric Villard}

\address{%
$^1$ \quad European Southern Observatory, Karl-Schwarzschildstr. 2, Garching D-85748, Germany
\\
$^2$ \quad Max Planck Institute for Astrophysics, Karl-Schwarzschild-Str.1, Garching D-85748, Germany

}
\corres{lukasz.tychoniec@eso.org}

\firstnote{Submitted to International Workshop on Bayesian Inference and Maximum Entropy Methods in Science and Engineering, IHP, Paris, July 18-22, 2022.} 

\abstract{ The Atacama Large Millimeter/submillimeter Array (ALMA) is currently revolutionizing observational astrophysics. The aperture synthesis technique provides angular resolution otherwise unachievable with the conventional single-aperture telescope. 
However, recovering the image from the inherently undersampled data is a challenging task. The \texttt{clean} algorithm \cite{Hoegbom1974} has proven successful and reliable and is commonly used in imaging the interferometric observations. It is not, however,  free of limitations. 
Point-source assumption, central to the  \texttt{clean} is not optimal for the extended structures of molecular gas recovered by ALMA. Additionally, negative fluxes recovered with  \texttt{clean} are not physical. This begs to search for alternatives that would be better suited for specific science cases.
We present the recent developments in imaging ALMA data using Bayesian inference techniques, namely the \texttt{resolve} algorithm \cite{Junklewitz.Bell.ea2015}. This algorithm, based on information field theory \cite{Ensslin2013}, has been already successfully applied to image the Very Large Array data \cite{Arras.Bester.ea2021}.
We compare the capability of both  \texttt{clean} and  \texttt{resolve} to recover known sky signal, convoluted with the simulator of ALMA observation data  and we investigate the problem with a set of actual ALMA observations.
}


\begin{document}

\section{Introduction}
\label{sec:Introduction}

\subsection{Aperture synthesis}
The Atacama Large Millimeter/submillimeter Array (ALMA) is revolutionizing  observational astrophysics. With its 66 antennas located on the Atacama desert it provided the sharpest ever images of the submillimeter sky, for example, images of the protoplanetary disks at 1 au resolution \cite{Andrews.Wilner.ea2016}. In order to obtain such a resolution at a distance to a nearby star-forming region at 1.3 mm a telescope diameter of $\sim 15$ km are needed. Since the construction challenges of such an antenna, especially if one would like to make it steerable, are far beyond current technical capabilities, in radio astronomy domain we often turn to aperture synthesis techniques, where instead of a single dish, a combination of smaller antennas is used, and with interference of signal between each antennas a resolution compared to the a telescope of a size of the greatest distance between the two antennas in an array (i.e. baseline) is achieved. 

This does not come without a cost: sampling the baselines is never complete compared with a single dish telescopes. This means we do not receive complete information at all baselines and therefore to create an image of the sky we are operating with missing information. 

A direct measurement of an interferometer is the interference pattern between two given antennas. This pattern is related to the sky brightness observed by the antennas. The recorded complex value called {\it visibility} is then a Fourier transform of the sky brightness, the quantity which observations aim to recover. Therefore, a simplified imaging process consists of a (reverse) Fourier transformation of the measured visibilities (while the non-measured visibilities are set to 0) to obtain first approximation of the sky brightness, the so-called {\it dirty image} \citep{Jackson2008,Thompson.Moran.ea2001}. Once this is achieved, it becomes apparent that even with modern interferometer like ALMA with many baselines sampling the so-called UV plane we achieve a rather poor quality image. Refining this image is the topic of this work.

\subsection{CLEAN as fa standard approach to the imaging of the interferometric data}

A standard technique to improve the image quality of the interferometric observations has been developed by Hogbom in 1974 \cite{Hoegbom1974} and is called \texttt{clean}. \texttt{Clean} makes use of the well-defined point-spread function of a given antenna configuration. The algorithm identifies the point sources in the initial dirty image, point sources are then approximated with a delta function, convolved it with a dirty beam (i.e. the assumed pattern that the point-source would create on the dirty image), scales it with brightness of the suspected point source and subtracts the point-source patter from the dirty image. This is an iterative process which ideally ends when all point sources are removed from the dirty image so that it consists only of noise. \citep{Jackson2008}

One issue of \texttt{clean} that can be easily identified is in the assumption that the sky is composed of point sources. This results in \texttt{clean} struggling with imaging of extended sky brightness structures. Several modifications to the original \texttt{clean} algorithm have been implemented to mitigate this issue, such as multi-scale clean which allows to set a point-source that would be a Gaussian rather than a delta function \cite{Junklewitz.Bell.ea2015,Cornwell2008}.

There are several steps in the process of cleaning which can be modified to improve the process. Masking is a method to restrict the area where the algorithm will look for point sources to a selected region on the sky. Weighting allows to attribute different weights to different u,v scales, allowing for a trade-off between resolution and sensitivity. 

Although \texttt{clean} became a gold standard of the interferometric data imaging, producing many stunning images of the astronomical objects its limitation beg to seek alternatives, especially in cases where its assumptions are not met.


\subsection{\texttt{resolve} algorithm and IFT}
Imaging of the interferometric data can be presented as an inference problem, since we operate on an incomplete measurement problem, trying to find the true sky emission from the received data. Radio Extended SOurces Lognormal deconVolution Estimator \texttt{resolve} \footnote{https://gitlab.mpcdf.mpg.de/ift/resolve} \cite{Junklewitz.Bell.ea2015,Arras.Knollmueller.ea2018} is designed in the Information Field Theory (IFT) framework \citep{Ensslin.Frommert.ea2009}. IFT enables to use Bayesian inference methods in the context of mathematical framework of field theory. This is well fitted to the issue of imaging the sky brightness. IFT algorithms are implemented in \texttt{resolve} through Python package NIFTy \footnote{https://gitlab.mpcdf.mpg.de/ift/NIFTy} \cite{Selig.Bell.ea2013,Arras.Baltac.ea2019}.

The measurement equation can be presented as: $d = R(s) + n$, where $R$ is a response of the instrument to the original physical signal ($s$), and $n$ is noise. In the IFT framework, obtained data $d$ is analyzed in order to find the most probable form of $s$, which is signal from the field, in this case, the sky. This takes the form of Bayes' theorem:
\begin{equation}
     P(s|d) =\frac{P(d|s)P(s)}{P(d)},
\end{equation}
$P(d|s)$ - likelihood that a given data (d) has been produced from a signal (s), $P(s)$ - is a prior knowledge about the signal, and P(d) is a normalization factor.

\texttt{Resolve} is developed in order to optimize imaging of extended and diffuse radio sources and to provide reliable noise estimation \cite{Junklewitz.Bell.ea2015}. It has been successfully used for example in imaging of Cygnus A \citep{Arras.Bester.ea2021} observed with the VLA and the M87* black hole environment with VLBI \citep{Arras.Frank.ea2022}. In this work we want to explore \texttt{resolve} capabilities for ALMA interferometer, specifically compared with the most commonly used \texttt{clean} algorithm.



\section{Test datasets} 
In this section we present data used to test interferometric imaging methods: \texttt{resolve} and \texttt{clean}. We use a simple simulated dataset in order to have a  complete control over the input information, as well as the real ALMA observations.
\subsection{Simulated data}
\label{sec:simulated_data}

In this section, we describe the preparation of the simulated data. The major advantage in using simulated datasets to analyze the imaging procedures is a complete control of the input parameters. We control perfectly the shape and brightness of the sources, as well as configuration and behaviour of the telescope array. On the downside of this approach is the difficulty to realistically model the noise acquired during observation and simplistic assumptions about the sky brightness.

\begin{figure}
    \includegraphics[width=0.5\textwidth]{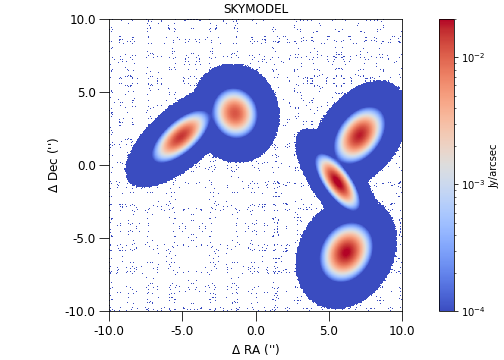}
    \includegraphics[width=0.45\textwidth]{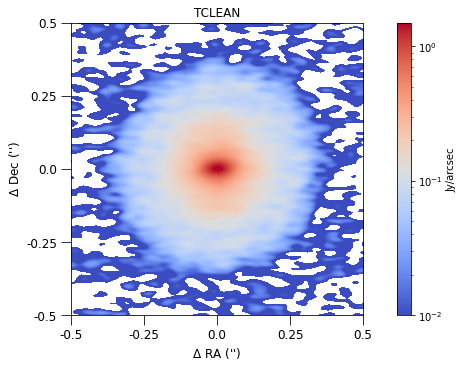}
    \caption{{\it Left:} Sky model which serves as an input for creating simulated ALMA Measurement Set. {\it Right:} Best \texttt{tclean} image of Sz114 protoplanetary disk from ALMA observations.}
    \label{fig:simalma_skymodel}
\end{figure}

We generate a simple 2D array consisting of five Gaussian components of different brightness, size and position angle. This model is input into the Common Astronomy Software Applications package (CASA; \cite{McMullin.Waters.ea2007}) task \texttt{simalma}. This task converts image to the sky model, imposing physical properties to the sky, such as spherical coordinates, pixel dimensions, field-of-view and brightness in physical units. Fig. \ref{fig:simalma_skymodel} (left) shows the sky model generated with \texttt{simalma} task in CASA. 
Afterwards, the task is simulating observations of the given sky model with ALMA observatory. For this specific case we simulate observations with ALMA configuration C-3 at 230 GHz (ALMA band 6), which results in effective resolution of 0.7", determined approximately by largest available baseline (distance between two antennas). In the case of C-3 configuration this is $\sim$ 500 m.

The \texttt{simalma} task returns a calibrated Measurement Set (MS), that consists of complex visibilities. Those visibilities are Fourier transform of the sky brightness, therefore reverse Fourier transformation provides a dirty image of the observed sky. Further on we describe the process of imaging with \texttt{resolve} and \texttt{tclean}.

\subsection{ALMA data}

Here we describe archival ALMA observations used to test application of \texttt{resolve} on real-life example. It is important to compare it with a well-understood case, in which the calibration and imaging examples with other tools are already available. We select a protoplanetary disk Sz114 observed within DSHARP ALMA Large Program \citep{Andrews.Huang.ea2018a}. This was a milestone program for ALMA observatory, showcasing the richness of substructures within disks, often associated with ongoing planet formation. It is therefore especially interesting to search for innovative methods of imaging those data in order to verify current conclusions, as well as open avenue for new studies.

The Sz114 disk shows relatively smooth structure compared with other extreme cases and therefore it serves well as a test case to try to identify underlying structure with \texttt{resolve}. Fig. \ref{fig:simalma_skymodel} (right) presents image of the disk obtained with \texttt{tclean} as delivered by the Large Program team \citep{Andrews.Huang.ea2018a}.

\subsection{Imaging of the simulated dataset}

First, in order to image the simulated measurement set, we use \texttt{tclean} task in CASA software which implements CLEAN algorithm as in Hogbom 1974 \cite{Hoegbom1974}. We ran \texttt{tclean} without any constraint on where to look for point sources in the image (i.e. without any masking), for 20000 iterations, or until threshold of 0.3 mJy/beam was reached. Pixel size of the reconstructed image is set to 0.1" and image size is set to 512$\times$512 pixels.

\begin{figure}

        \includegraphics[width=0.5\textwidth]{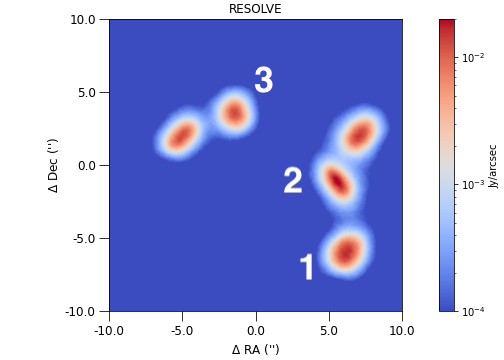}
        \includegraphics[width=0.5\textwidth]{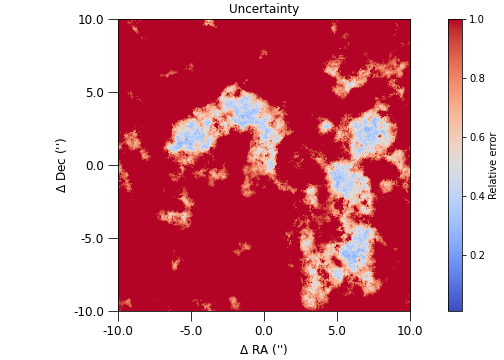}
                \includegraphics[width=0.5\textwidth]{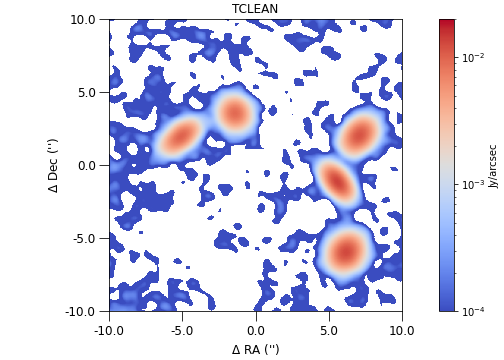}
        \includegraphics[width=0.5\textwidth]{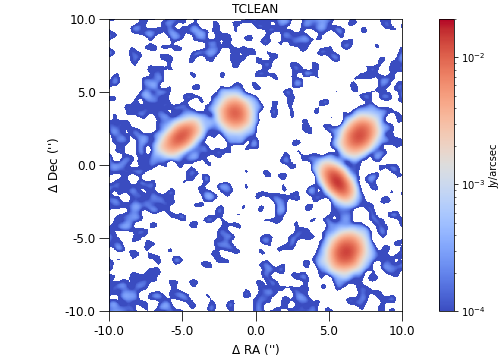}
    \label{fig:simalma_res_cl}
        \caption{Comparison of the imaging of simulated ALMA observations with  \texttt{resolve}  and  \texttt{tclean}. {\it Top:} \texttt{resolve} image (left) with associated uncertainty map. {\it Bottom:} two \texttt{tclean} images, left: with natural weighting; right: with Briggs weighting and robust = 0.5.}
\end{figure}

This threshold is set based on the earlier, quick \texttt{tclean}, so that the algorithm does not attempt to find sources from the residual image consisting purely of noise. In the default settings, weighting of baselines is set to 'natural', which means it associate the baseline with weight proportional to the sampling density (i.e. the most covered baselines have the highest weight). Since there is much more baselines sampling the larger scales, this results in putting more weight to lower resolution, which results in achieving lower resolution than the sky model image. Therefore we also attempt an imaging with Briggs weighting with robust parameter 0.5, which moves the balance of weighting toward longer baselines increasing the resolution but decreasing signal-to-noise ratio. The result of the  \texttt{tclean} run is presented in Fig. \ref{fig:simalma_res_cl} (bottom).

For \texttt{resolve} image of the simulated data we run 30 iterations, using Newton optimizer. We did not make any assumptions on the presence of the point sources in the data. Parameters of the \texttt{resolve} run are summarized in Table \ref{tab:tab_resolve}. For detailed explanation of the parameters see Arras et al. \cite{Arras.Bester.ea2021}.
\vspace{1cm}

\begin{table}
\caption{Parameters of the \texttt{resolve} run on simulated dataset (left) and Sz114 ALMA data (right)}

\centering
\begin{minipage}{0.45\linewidth}
\label{tab:1}
\begin{tabular}{l c c c c c c  r r c c c cc}        
\hline\hline                 
Parameter  & Mean\\
\hline                                   
Offset &  26 \\
Zero mode  & 1$\pm0.1$ \\
Fluctuations &  5$\pm1$\\
Power spectrum slope & -2$\pm0.5$\\
Flexibility & 1.2$\pm0.4$\\
Asperity & 0.2$\pm$0.2\\

\hline                                   

\end{tabular}
\end{minipage}
\begin{minipage}{0.45\linewidth}
\label{tab:2}
\begin{tabular}{l c c c c c c  r r c c c cc}        
\hline\hline                 
Parameter  & Mean\\
\hline                                   
Offset &  20 \\
Zero mode  & 1$\pm0.2$ \\
Fluctuations &  3$\pm1$\\
Power spectrum slope & -4$\pm2$\\
Flexibility & 4$\pm0.8$\\
Asperity & 2$\pm$0.8\\

\hline                                   

\end{tabular}
\end{minipage}

    \label{tab:tab_resolve}
\end{table}

\subsection{Imaging of the real ALMA observations}

For the Sz114 protoplanetary disk observed within DSHAPR ALMA Large Program \cite{Andrews.Huang.ea2018a} we have created an image from a single spectral window, in order for a direct comparison with \texttt{resolve}, which currently does not have a capability to make images of multiple spectral windows. From the publicly available MS file we extracted spectral window 9 with \texttt{split} task in CASA and binned it into a single channel. 

For this \texttt{tclean} run we implemented both standard and multiscale imaging in order to compare the outcomes, since the observed disk presents large variety of spatial scales. We create images with 0.005" and 1024$\times$1024 pixels and run 20000 iterations up to noise threshold of 0.05 mJy is reached. In case of multiscale clean we specified three scales: at 0, 7, and 28 pixels, which means the \texttt{tclean} algorithm iterate in order to find three types of sources in the data: either a point source, which corresponds to scale 0, and extended components with Gaussian shape of 7 and 28 pixels of FWHM. Results of the imaging are presented in Fig. \ref{fig:Sz114_res_cl} (bottom).

A \texttt{resolve} imaging followed the same procedure as in the simulated data, however, since the weights were obtained with from real observation, this results in more realistic treatment of the uncertainties. Parameters of this run are shown in Tab. \ref{tab:tab_resolve}.

\section{Comparison to \texttt{tclean} and \texttt{resolve}}
In this initial study of performance of the \texttt{resolve} algorithm in operating with ALMA data, we focus on key aspects of the imaging process: flux recovery -- how much flux observed on the sky is recovered by the imaging process and how accurate it is; image quality -- effective resolution and dynamic range.

\subsection{Simulations}

\begin{figure}

        \includegraphics[width=0.5\textwidth]{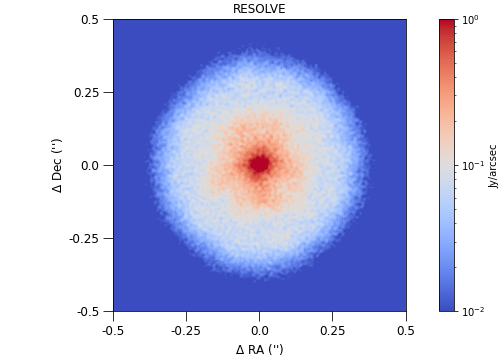}
         \includegraphics[width=0.5\textwidth]{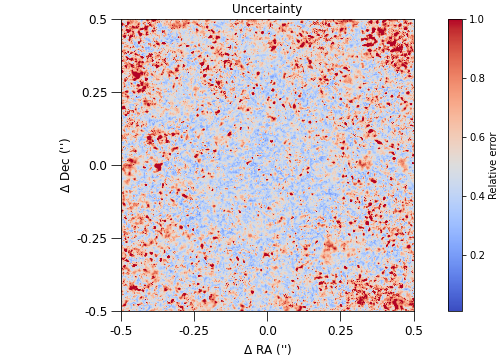}
          \includegraphics[width=0.5\textwidth]{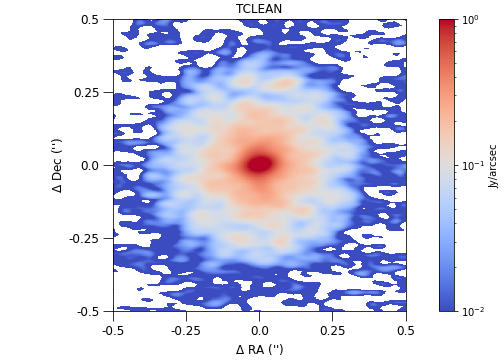}
          \includegraphics[width=0.5\textwidth]{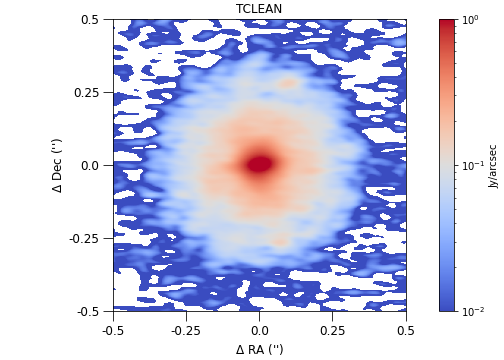}
    \caption{Comparison of the imaging of Sz114 ALMA observations with resolve and tclean. {\it Top:} \texttt{resolve} image (left) with associated relative error map. {\it Bottom:} two \texttt{tclean} images, left: with standard cleaning algorithm, right: with multi-scale algorithm}
    \label{fig:Sz114_res_cl}
\end{figure}

We compare the quality of \texttt{resolve} and \texttt{tclean} imaging of the simple simulated ALMA observation, as described in Section \ref{sec:simulated_data}. Fig. \ref{fig:Sz114_res_cl} presents \texttt{resolve} image and associated uncertainty map, and \texttt{tclean} image with two different weighting schemes: natural and Briggs weighting with robust parameter of 0.5. 

We note that for \texttt{tclean} the imaging resulted with large areas with a negative flux, which do not occur in \texttt{resolve} - positive-only emission is one of the assumptions in the prior. On the uncertainty map produce by \texttt{resolve} we observe much higher confidence associated with the source positions, compared with positions where no significant emission was detected.

We measure integrated fluxes of each of the Gaussian components in both imaging methods as well as in the input sky model. Result is presented in Table \ref{tab:1}. We can note that \texttt{resolve} and \texttt{tclean} both underestimate the flux of the brightest source (i.e. the Gaussian associated with the largest integrated flux). In case of \texttt{tclean} this can be attributed to imperfect recovery of the signal from the point-spread function, but it is unclear why \texttt{resolve} underpredicts the flux. More extensive runs are needed to investigate this issue.

On the other hand 
\texttt{resolve} resolve reaches good accuracy on the other two measured components. We are able to provide much more reliable error bars on \texttt{resolve} by taking the standard deviation of the measured value from all the samples in the final iteration.

\begin{table}

\begin{minipage}{0.5\textwidth}
\caption{Peak flux (in Jy arcsec$^{-2}$) of different components on the simulated image.}
\begin{tabular}{l l c c c c c  r r c c c cc}        
\hline\hline                 
Comp & model & \texttt{tclean} & \texttt{tclean}  & \texttt{resolve}\\
 & & natural  & Briggs &  &   \\
\hline                                   

1 &  34.11 & 24.67 & 24.96 & 26.68$\pm$5.76 \\
2 &  23.31 & 21.68 & 22.19 & 23.59$\pm$6.72 \\
3 &  16.69 & 16.31 & 16.57 & 17.82$\pm$3.86\\
\hline                                   
\label{tab:1}
\end{tabular}

\end{minipage}
\begin{minipage}{0.5\textwidth}

\caption{Peak (in Jy arcsec$^{-2}$) and integrated flux at given radii of the Sz114 disk.}
\begin{tabular}{l l c c c c c  r r c c c cc}        
\hline\hline                 
Radius & best & \texttt {tclean} & \texttt{tclean}  & \texttt{resolve}\\
 & & hogbom  & multiscale &  &   \\
\hline          
peak &  1.57 & 1.55 & 1.49 & 5.8$\pm$3.82 \\
0.06 &  8.63 & 8.85 & 8.95 & 9.48$\pm$1.07 \\
0.15 &  22.79 & 22.82 & 22.53 & 23.21$\pm$0.68 \\
0.35 &  47.45 & 47.07 & 47.24 & 47.24$\pm$2.79 \\
\hline                                   
\label{tab:2}
\end{tabular}

\end{minipage}

\end{table}

\subsection{ALMA data}

In the case of real ALMA data it is more difficult to assess total recovered flux since we do not have information on true flux. We compare the peak and total flux integrating over different areas of the disk \ref{tab:2}. First of all, it can be noted that all \texttt{tclean} images have comparable fluxes, therefore the multiscale did not affect significantly the flux measurements. 

In comparison between \texttt{resolve} and \texttt{tclean}, we can see that peak flux measured from \texttt{resolve} image is 3 times as high as the peak in \texttt{tclean} images. This can be purely gridding effect as the pixel scale of the \texttt{resolve} image is much smaller.

At the same time we note slightly comparable values of the integrated flux measured at from different radii of the disk.

At first sight it appears that \texttt{resolve} is able to create super-resolution image of the disk. We can use the uncertainty map to understand what confidence can be associated with those structures. We note on the uncertainty map that we typically achieve better than 20$\%$ confidence on the results but the map also reaches very low confidence in some areas of the map.



\section{Conclusions}

In this work we apply Bayesian inference and field theory in the framework of Information Field Theory (IFT) using the \texttt{resolve} algorithm, to create images from radio intereferometric observations obtained with ALMA array. The key conclusions are:

\begin{itemize}
    \item the imaging of simulated ALMA Measurement Set with \texttt{resolve} results in a  successful recovery of the flux of the different components in the simulated dataset.
    \item In one of the first attempts to apply \texttt{resolve} on real ALMA data we obtain a high-fidelity image of protoplanetary disk Sz 114, highlighting the potential of \texttt{resolve} to create super-resolution images.
    \item For both test cases we obtain a robust estimation of the uncertainties on the measured fluxes, which is one of the major advantages of \texttt{resolve} compared with \texttt{tclean}.
\end{itemize}

With those encouraging results we highlight further areas where exciting developments can be made: \texttt{resolve} can be used to create spectral cubes, using correlation between channels as additional prior information; create combined maps of different antenna configuration, a particularly challenging case for \texttt{tclean} algorithm and where accurate assessment of the confidence in obtained data is especially necessary. 

{\bf Acknowledgments:} This work made use of the following software: resolve \cite{Junklewitz.Bell.ea2015}, matplotlib \cite{Hunter2007}, NIFTy v.8 \cite{nifty}, astropy \cite{AstropyCollaboration.PriceWhelan.ea2018}, CASA \cite{McMullin.Waters.ea2007}, This research has made use of NASA’s Astrophysics Data System Bibliographic Services. This paper makes use of the following ALMA data: ADS/JAO.ALMA\#2016.1.00484.L. ALMA is a partnership of ESO (representing its member states), NSF (USA) and NINS (Japan), together with NRC (Canada), MOST and ASIAA (Taiwan), and KASI (Republic of Korea), in cooperation with the Republic of Chile. The Joint ALMA Observatory is operated by ESO, AUI/NRAO and NAOJ. 







\reftitle{References}

\bibliography{main}

\end{document}